\documentclass[sigconf]{acmart}

\AtBeginDocument{%
  \providecommand\BibTeX{{%
    \normalfont B\kern-0.5em{\scshape i\kern-0.25em b}\kern-0.8em\TeX}}}

\copyrightyear{2023}
\acmYear{2023}
\setcopyright{acmlicensed}\acmConference[PASC '23]{Platform for Advanced Scientific Computing Conference}{June 26--28, 2023}{Davos, Switzerland}
\acmBooktitle{Platform for Advanced Scientific Computing Conference (PASC '23), June 26--28, 2023, Davos, Switzerland}
\acmPrice{15.00}
\acmDOI{10.1145/3592979.3593417}
\acmISBN{979-8-4007-0190-0/23/06}

\usepackage{orcidlink}
\usepackage{color}
\usepackage{ulem}

\begin{document}

\title[Cornerstone]{Cornerstone: Octree Construction Algorithms for Scalable Particle Simulations}
\author{Sebastian Keller \orcidlink{0000-0003-3540-1405}}
 \email{sebastian.keller@cscs.ch}
 \affiliation{Swiss National Supercomputing Center (CSCS) \country{Switzerland}}
\author{Aur\'elien Cavelan \orcidlink{aurelien.cavelan@unibas.ch}}%
\email{aurelien.cavelan@unibas.ch}
\affiliation{%
University of Basel \country{Switzerland}
}%
\author{Rub\'en Cabezon \orcidlink{0000-0003-3546-3964}}%
\email{ruben.cabezon@unibas.ch}
\affiliation{%
University of Basel \country{Switzerland}
}%
\author{Lucio Mayer \orcidlink{0000-0002-7078-2074}}
 \email{lmayer@physik.uzh.ch}
\affiliation{%
University of Z\"urich \country{Switzerland}
}%
\author{Florina M. Ciorba \orcidlink{0000-0002-2773-4499}}
\email{florina.ciorba@unibas.ch}
\affiliation{%
University of Basel \country{Switzerland}
}%

\date{\today}

\begin{abstract}
This paper presents an octree construction method, called Cornerstone, that facilitates global domain decomposition and
interactions between particles in mesh-free numerical simulations.
Our method is based on algorithms developed for 3D computer graphics, which we
extend to distributed high performance computing (HPC) systems. 
Cornerstone yields global and locally essential octrees and is able to
operate on all levels of tree hierarchies in parallel.
The resulting octrees are suitable for supporting the computation of various kinds
of short and long range interactions in N-body methods, such as Barnes-Hut and the Fast Multipole Method (FMM).
While we provide a CPU implementation, Cornerstone may run entirely on GPUs.
This results in significantly faster tree construction compared to execution on CPUs
and serves as a powerful building block for the design of simulation codes
that move beyond an offloading approach, where only numerically intensive tasks are dispatched to GPUs.
With data residing exclusively in GPU memory, Cornerstone eliminates data movements between CPUs and GPUs.
As an example, we employ Cornerstone to generate locally essential octrees for a Barnes-Hut treecode
running on almost the full LUMI-G system with up to 8 trillion particles.
\end{abstract}

\maketitle

\section{Introduction}
\label{sec:introduction}

Octrees serve a variety of purposes in particle-based simulations.
With applications in Astrophysics and Cosmology \cite{Hernquist1987, WarrenSalmon1993, Pkdgrav3},
Smoothed Particle Hydrodynamics (SPH) \cite{Gingold1977, Lucy1977}, and Molecular Dynamics \cite{GromacsFMM},
they are typically employed to decompose domains, identify halo particle regions, solve the neighbor-search problem,
calculate global magnitudes efficiently \cite{Perego2014}, and support the analysis of simulation data \cite{Knollmann2009}.
If long-range interactions are involved, as is the case in electrostatics or gravity,
octrees also form the basis for algorithms like Barnes-Hut \cite{Barnes1986} and the Fast Multipole Method (FMM) \cite{Greengard1987}.
With ever increasing compute power ratios between GPUs and CPUs, the need to also perform tree construction and traversal
on the GPU is increasing, despite the usually lower numerical intensity compared to the particle force calculations.
With seemingly sequential dependencies between tree levels and memory layout concerns, efficient octree construction and traversal on GPUs remain a difficult problem. 

Nevertheless, there has been a lot of progress in recent years in the field of 3D computer
graphics where octrees form the first step in the construction of
bounding volume hierarchies for collision detection.
Several octree construction algorithms \cite{Karras2012, Apetrei2014}
operate in parallel on all levels of the tree hierarchy and store data compactly in linear buffers.
As a first step, these algorithms rely on the computation of sorted Morton codes \cite{Morton1966} for the
3D objects present in the scene,
which are then treated as the leaves of a binary radix tree.
Hence, if we were to apply these algorithms directly to the particles in a numerical simulation,
the resulting binary radix tree would resolve every particle in a separate leaf node.
When calculating physical particle-particle interactions, however, we have the option of reducing the
number of tree nodes and allowing multiple particles per leaf node.
With increasing number of particles per leaf, the overhead due to processing tree nodes decreases while the
overhead due to decreasing search efficiency for neighbor particles increases.
These two opposite trends lead to an optimum balance that
must be determined empirically by varying the maximum number of particles permitted per leaf node;
this is commonly referred to as $N_{crit}$ \cite{Barnes1990}.

In this work, we present a new octree construction method that combines the favorable property of operating
on all levels of the tree in parallel on GPUs with the option of choosing $N_\mathrm{crit} > 1$.
Starting from the aforementioned methods developed for 3D computer graphics, we achieve
this by aggregating the space filling curve (SFC) \cite{Peano1890} keys of particles,
into a histogram before generating the internal tree nodes.
The placement and size of the histogram bins are themselves defined by SFC keys and
are subject to certain constraints that cause each bin to correspond to a leaf node of an octree.
All bins together cover the entire SFC, and thus each SFC key in the histogram delimits
the start and end of an octree leaf node, which is why we call this histogram cornerstone array.
Our method works for the Morton Z-curve as well as a 3D version of the
Hilbert curve \cite{Hilbert1891, Haverkort2011}. We employ the latter in our simulations
due to its superior locality preserving properties compared to the Morton Z-curve.

We then apply the cornerstone array to the computation
of locally essential trees (LET) \cite{Warren1992} in the context of distributed numerical particle simulations.
A novel feature of our LET implementation is that the determination of LET branches to be
exchanged with remote subdomains is carried out on GPUs, as is the subsequent communication. 

Finally, as a proof of concept, we present a Barnes-Hut treecode implementation that traverses the LET
generated by our method. It is performance-portable between AMD and NVIDIA GPUs and able to scale to
trillions of particles.

In the following, Sec.\ \ref{sec:sfc} introduces relevant properties of SFCs and their connections to octrees.
Sec.\ \ref{sec:leaftree} describes our iterative method for generating octree leaf nodes with particle counts
bounded by $N_\mathrm{crit}$.
while Sec.\ \ref{sec:linkedtree} illustrates how the internal part of the octree can be computed.
The extension to distributed octrees is discussed in Secs.\ \ref{sec:distributedtree} and\ \ref{sec:essentialtree} before
we discuss results and future work in Sec.\ \ref{sec:results}.

\section{Related work}

In the context of N-body simulations, Burtscher and Pingali \cite{Burtscher2011}, shortly followed by B\'edorf et al.,
\cite{Bonsai2012} were the first to generate octrees on GPUs as the basis for Barnes-Hut treecodes. In both approaches,
the octree was constructed level by level on a single GPU.
B\'edorf et al. later extended their method \cite{Bonsai2014} to distributed HPC systems with GPU accelerators
and combined SFC-based domain decomposition with the LET method, handling the non-local parts of generating the LET
on the CPU.

With the aim of improving the speed of bounding volume hierarchy generation for applications in
3D computer graphics, Karras \cite{Karras2012} was the first to present an algorithm for binary radix tree construction
that generated all levels of the tree in parallel and used it as a building block to generate octrees as well.
Apetrei \cite{Apetrei2014} further refined the algorithm by replacing binary searches with atomic operations.
A shared trait between both approaches is that the resulting binary radix or octrees resolve each 3D object in a separate leaf cell.

\section{Space filling curves and octrees}

\label{sec:sfc}

Space filling curves (SFCs) are continuous functions that map the unit interval into an $n$-dimensional hypercube.
In conjunction with three-dimensional particle simulations, we are interested in discretized versions of SFCs that map the
key space $[0 \ldots 2^{3L})$ onto a grid of
3D integer coordinates $[0 \ldots 2^L) \times [0 \ldots 2^L) \times [0 \ldots 2^L)$.
In 3D, the number of bits required to store an SFC key or a point on the grid equals $3L$.
Since current computer architectures have instructions for either 32- or 64-bit integers, reasonable choices for $L$ are $10$ or $21$,
or -- depending on accuracy requirements -- multiples thereof.

The utility of certain SFCs for numerical simulations stems from their relation to octrees.
If we express a key $k$ of the Morton Z-curve with $3L$ bits as a sequence of $L$ octal digits,
\begin{equation}
    k = k_1 k_2 \ldots k_L,
\end{equation}
Warren and Salmon found \cite{WarrenSalmon1993} that if the first octal digit $l_1$ of another key $l$ matches $k_1$,
then $k$ and $l$ decode into 3D coordinates that lie in the same octant of the root octree node.
And by induction: if the first $i$ octal digits of keys $k$ and $l$ match, then there exists an octree node at the
$i$-th division level that contains the decoded 3D coordinates of both keys. 
Equivalently, an octree node at the $i$-th division level contains the 3D coordinates
that encode into the key range $[k \ldots k+8^{L-i})$ for some unique $k$ with $k \,\, \mathrm{mod} \,\, 8^{L-i} = 0$.
Consequently, the number of octal digits $L$ in the SFC key is equal to the octree depth that the key is able to resolve.
More generally, this correspondence between octal digits of the key and nodes of an octree applies to any type of SFC
that traverses a cube octant by octant, with the Hilbert curv  as a further example.
The encoding and decoding of 3D grid points into Hilbert keys is computationally expensive compared to the simpler Morton Z-curve,
but in contrast to the latter, any continuous segment of the curve is mapped to a compact 3D volume,
while an interval of Morton keys may correspond to disconnected 3D volumes.
In distributed simulations, the smaller surfaces of subdomains defined as segments of the Hilbert curve require less communication,
outweighing the higher computational cost of key encoding compared to the Morton Z-curve \cite{Bonsai2014}.

\section{Leaf nodes of balanced octrees}
\label{sec:leaftree}

As described in the previous section, arrays of space filling curve (SFC) keys can be
used to represent the leaves of arbitrary octrees.
Let $\mathbf{K}$ be an array of SFC keys of length $n_l+1$ subject to the following constraints:
\begin{eqnarray}
    &&\mathbf{K}_0 = 0, \nonumber \\
    &&\mathbf{K}_{n_l} = 8^L, \nonumber \\
    &&\mathbf{K}_i < \mathbf{K}_j, \quad i < j, \quad \text{and} \nonumber \\
    &&\mathbf{K}_{i+1} - \mathbf{K}_i = \delta_i = 8^l, \quad l \in \mathbb{N}_0, \,  l \leq L.
    \label{eq:cstone}
\end{eqnarray}
Then, $\mathbf{K}$, which we refer to as the cornerstone array, uniquely defines an octree with $n_l$
leaves and a maximum depth of $L$. In this notation and ordering,
the $i$-th leaf covers the key range $[\mathbf{K}_i, \mathbf{K}_{i+1})$
and its division level is $L - \log_8{\delta_i}.$
Given an ensemble of real-valued 3D particle coordinates, we want to calculate $\mathbf{K}$, such
that the leaf particle counts remain below a predefined threshold, i.e. $\mathbf{N}_i \leq N_\mathrm{crit}$,
while the particle counts of internal nodes exceed $N_\mathrm{crit}$.
If these conditions are satisfied, we say that $\mathbf{K}$ is balanced, which is a requirement for efficient
particle neighbor searches. Note that since here the term \textit{balanced} refers to a property of the leaf nodes
(the number of contained particles) rather than to a property of the tree itself, its depth can vary locally.
The variable $N_\mathrm{crit}$ is a performance tuning parameter in this context
whose optimal value has to be determined empirically. Commonly, $N_\mathrm{crit}$ is chosen in the range 16 to 64
\cite{Barnes1990, Yokota2013, Bonsai2014}.

Let $\mathbf{P}$ be the sorted array of particle SFC keys obtained by encoding the particle coordinates into
SFC keys followed by a radix-sort.
We define two operations:
\begin{eqnarray}
    &&f: (\mathbf{P}, \mathbf{K}) \mapsto \mathbf{N\phantom{'}} \quad (\text{keys histogram}) \label{eq:histogram}\\
    &&g: (\mathbf{K}, \mathbf{N}) \mapsto \mathbf{K'} \quad (\text{rebalancing}) \label{eq:rebalance}
\end{eqnarray}
The $f$ operation counts the number of particle keys in each leaf node, or
in other words, $\mathbf{N}$ is the histogram of particle SFC keys $\mathbf{P}$ for
the $n_l$ bins defined by consecutive keys in $\mathbf{K}$.
With $\mathbf{P}$ sorted, the particle count of the $i-th$ bin delimited by $\mathbf{K}_i$ and $\mathbf{K}_{i+1}$
can be determined by locating $\mathbf{K}_i$ and $\mathbf{K}_{i+1}$ in $\mathbf{P}$ with binary searches.
The particle count of the bin is then given by the difference of the two delimiter positions.
The $g$ operation rebalances a given cornerstone array based on particle counts where
each leaf node can either remain unchanged, be subdivided or be merged. The merging of leaf nodes
is achieved by replacing a group of 8 sibling nodes with their common parent node.

We further subdivide $g = g_3 \circ g_2 \circ g_1$ into three steps:
\begin{alignat}{3}
&g_1: (\mathbf{K}, \mathbf{N}) \mapsto \mathbf{O}, \quad &&(\text{rebalancing decision}) \label{eq:g1}\\
&g_2: \mathbf{O} \mapsto \mathbf{O'}                     &&(\text{exclusive prefix sum}) \label{eq:g2}\\
&g_3: (\mathbf{K}, \mathbf{O'}) \mapsto \mathbf{K'}      &&(\text{node rebalancing}) \label{eq:g3}
\end{alignat}
First, $g_1$ computes an auxiliary array $\mathbf{O}$ that encodes the rebalance operation that is to be performed
on each leaf node, which can be either removal, subdivision, or no change of the leaf node.
These sub-operations are encoded as follows:
\begin{itemize}
\item $\mathbf{O}_i = 8$ if $\mathbf{N}_i > N_\mathrm{crit}$, indicating that leaf $\mathbf{K}_i$ is to be subdivided.
\item $\mathbf{O}_i = 0$ if
  \begin{itemize}
    \item $\mathbf{K}_i$ is not the first octant.
    \item $\mathbf{K}_i$ and its other 7 sibling octants are next to each other, i.e. when
            $$    \mathbf{K}_{i-j} + 8 \delta_i = \mathbf{K}_{i-j+8},$$
            with    
            $ j = \frac{\mathbf{K}_i \, \text{mod} \, 8\delta_i}{ \delta_i } \in [0 \ldots 7] $.
          This is only possible if all 8 octants are leaves.
    \item The combined particle count of these 8 octants is smaller than $N_\mathrm{crit}$.
  \end{itemize}
If all three conditions are fulfilled $\mathbf{K}_i$ will not be transferred to $\mathbf{K'}$, resulting in
the replacement of a group of 8 sibling nodes with their parent node.
\item $\mathbf{O}_i = 1$ otherwise, leaving the leaf node unchanged.
\end{itemize}
Subsequently, $\mathbf{O'}$ is obtained through $g_2$ by performing an exclusive prefix sum on $\mathbf{O}$.
By construction, $\mathbf{O'}_i$ is equal to the index of the old node $\mathbf{K}_i$ in the rebalanced
node array $\mathbf{K'}$, and $\mathbf{O'}_{n_l}$ is equal to the total number of rebalanced leaf nodes in $\mathbf{K'}$.
The rebalance step $g_3$ therefore constructs $\mathbf{K'}$ as follows:
\begin{alignat}{3}
\mathbf{K'}_{\mathbf{O'}_i} &= \mathbf{K}_i \quad &&\text{if} \quad \mathbf{O}_{i} = 1, \label{eq:rebalancecopy}\\
\mathbf{K'}_{\mathbf{O'}_i + j} &= \mathbf{K}_i + j \cdot \delta_i / 8,  j = [0\ldots 7]
    \quad &&\text{if} \quad \mathbf{O}_{i} = 8. \label{eq:rebalancesplit}
\end{alignat}

In Eq. \eqref{eq:rebalancecopy}, the SFC key at $\mathbf{K}_i$ is copied to its new location
$\mathbf{O'}_i$ in the rebalanced array $\mathbf{K'}$  while in Eq. \eqref{eq:rebalancesplit},
$\mathbf{K}_i$ is subdivided by inserting into $\mathbf{K'}$ the 8 SFC keys of its child nodes
that divide the original key range $[\mathbf{K}_i \ldots \mathbf{K}_{i+1})$ into 8 sub-ranges of equal length.
\begin{figure}
\vspace{1cm}
\includegraphics[width=\linewidth]{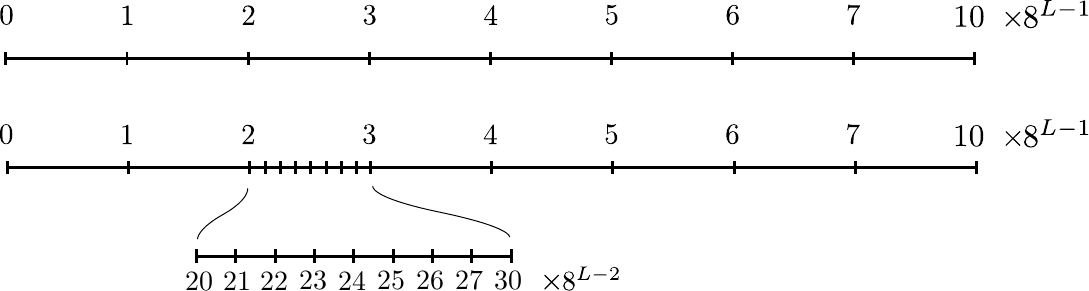}
\caption{Example of a cornerstone array undergoing refinement. The third octant in the top panel is replaced by its 8 child
nodes, resulting in an array that corresponds to the octree shown in Fig.\ \ref{fig:octree}.
SFC keys are labeled in octal representation.}
\label{fig:cstone_array}
\end{figure}
Fig.\ \ref{fig:cstone_array} shows an example where one of the octants of the root node is further subdivided by inserting
the SFC keys of its children into the cornerstone array.

By applying $f$ and $g$ in alternating fashion, we can construct the leaf nodes of a balanced octree, i.e. one whose
leaf node particle counts remain below or equal to $N_\mathrm{crit}$, while the particle counts of internal nodes
are greater than $N_\mathrm{crit}$. 
Starting with the root node, $\mathbf{K} = \lbrace 0, 8^L \rbrace$,
alternating application of $f$ and $g$ then yields the balanced cornerstone array $\mathbf{K}'$.
We say that convergence is reached when $\mathbf{O}_i = 1, \, \forall i$,
i.e. when $g_1$ reports that no leaf nodes need to be merged or subdivided anymore, which will
take at most $L$ repeated invocations of $f$ and $g$.
When computing a cornerstone array by starting from just the root node, our method thus offers no
advantage over a conventional top-down level-sequential octree construction approach.
The utility of our method arises when octrees are built repeatedly for multiple steps with particle positions
only changing by a small amount between each step. This applies, for example, to particle simulations
that are performed for several time-steps. In this situation, we can maintain a cornerstone array balanced
by performing a single histogram count and rebalance update on the cornerstone array of the previous step.
Implicitly, $\mathbf{K}$ contains the size and location of
internal octree nodes as well, but traversal and storing additional node properties is not possible in a straightforward manner. 
We describe the explicit construction of internal nodes in section\ \ref{sec:linkedtree}.

In terms of computational building blocks, the construction of cornerstone arrays relies on radix sort for the particle SFC keys
and on and prefix sums for rebalancing, which are available from
parallel libraries for both CPU and GPU architectures, such as the Thrust library \cite{Thrust}.
The remaining operations, SFC key generation from 3D particle coordinates, $f$, $g_1$, and $g_3$ consist of purely independent operations on array elements.
Since there is no need for synchronization or communication, these operations are simple to implement in parallel with
directive-based (e.g., OpenMP, OpenACC) as well as accelerator-specific programming paradigms (e.g., CUDA, HIP).

\section{Fully linked octrees}
\label{sec:linkedtree}

Although cornerstone arrays implicitly contain the information of the full
octree, they are not traversable because parent-child node relationships are not stored.
To efficiently generate the missing internal nodes, we will adapt the algorithms by Karras \cite{Karras2012}
and Apetrei \cite{Apetrei2014}. Both originate in the computer graphics community and may be used for
3D collision detection via the generation of bounding volume hierarchies, or for ray tracing.

Either approach relies on the computation of sorted SFC keys for
the objects present in a 3D scene as a first step. 
Subsequently, the SFC keys of the objects are viewed as the leaves of a binary radix tree whose internal nodes can be generated in parallel.
A sequence of $N$ sorted object SFC keys leads to a hierarchy of $N-1$
internal binary radix nodes. Each binary radix node is associated with a key that corresponds to the bit-string
of the longest common prefix of all the keys that it covers, i.e. the root node has a key of length 0, and
its two children have keys of length 1 consisting of either a 0 or a 1-bit. While the ordering of the $N-1$ binary radix
nodes differs between the two approaches of Karras and Apetrei, either method defines a specific node layout that allows
the determination of the node bit-string keys as well as parent-child relationships of the $N-1$ internal binary radix nodes
in a data-parallel fashion. The resulting binary tree is compact in the sense that it does not contain
any empty nodes or nodes with only a single child, which is a favorable property when $N_\mathrm{crit} = 1$.
As Karras mentions, it can be converted to an octree by extracting the binary nodes whose key lengths are divisible by 3.
In practice, this conversion may be implemented in the same spirit as Eqs. \eqref{eq:g1} - \eqref{eq:g3}:
First, for each binary node, a value of 0 or 1 is stored in the $\mathbf{O}$ array depending on whether the node key length
is divisible by 3 or not. After a prefix sum of $\mathbf{O}$, the total number of octree nodes is then known and
the relevant binary nodes may be extracted in parallel.

Since the cornerstone arrays contain SFC keys as well, we may provide them instead of the particle SFC keys as inputs
to the binary radix tree construction. 
This allows us to take advantage of the fully parallel construction of internal nodes
as well as adjusting $N_\mathrm{crit}$ to our needs (constructing the binary radix tree from the particle SFC keys
would yield a tree with $N_\mathrm{crit} = 1$).
If we wanted to preserve the compactness property, we could now eliminate any empty buckets from $\mathbf{K}$ and
construct the internal part of the octree as described above. But as we're interested in the case $N_\mathrm{crit} > 1$,
we are going to explore a second option where we do not eliminate any empty buckets from $\mathbf{K}$, which is equivalent
to mandating that each internal octree node have exactly 8 children. In this case, we can derive an analytical mapping
between the keys in $\mathbf{K}$ and the internal octree nodes that they implicitly contain that allows us to
bypass the construction of an intermediate binary radix tree. The resulting octree will contain a small fraction
of empty nodes, depending on the value of $N_\mathrm{crit}$ and the distribution of particles, but for $N_\mathrm{crit} = 16 - 64$
we have found that the increase in construction speed outweighs the penalty of a slightly larger tree, even for highly
clustered distributions.

The data format for the complete octree consists of the following arrays of integer types:
\begin{enumerate}
    \item $\mathbf{NK}[n]$ in $[0, 2^{3L})$, the node SFC keys
    \item $\mathbf{CO}[n]$ in $[1, n)$, the connectivity array
    \item $\mathbf{LO}[L + 2]$ in $[1, n)$, the tree level offsets
\end{enumerate}
where $n = n_i + n_l$ is the total number of tree nodes, with $n_i$ as the number of internal nodes and $n_l$ as the number of leaves.
Since the number of children per node is either 0 or 8, $n_i = (n_l - 1) / 7$.
In brackets for each array, we first specify its length, followed by the range of values that its elements may assume.
The first array contains the SFC keys of each node in the placeholder bit format defined in Ref.\ \cite{WarrenSalmon1993},
which allows encoding the SFC range covered by the node with a single key.
As the SFC keys can be decoded into 3D coordinates, $\mathbf{NK}$ stores the 3D grid coordinates of each node.
The connectivity of the tree is contained in $\mathbf{CO}$ which stores the index of the first child.
The range of child indices of node $i$ is therefore given by $[\mathbf{CO}[i], \mathbf{CO}[i] + 8)$. 
If $\mathbf{CO}[i] = 0$, index $i$ corresponds to a leaf.
Lastly, the $i$-th element of $\mathbf{LO}$ contains the index of the first node at the $i$-th octree level where $L$ is the
maximum octree depth that the keys contained in $\mathbf{NK}$ are able to encode. Since we include the root node at level zero,
$L+1$ entries are required to store the offset of each level plus an additional element to mark the end of the last level.
\begin{figure}
\vspace{1cm}
\includegraphics[width=0.9\linewidth]{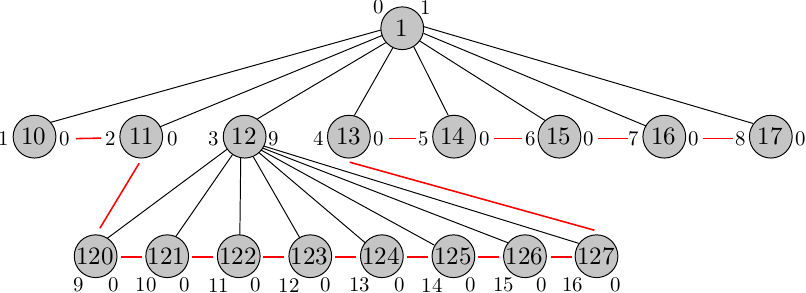}
\caption{A linked octree with node SFC keys in placeholder-bit format contained in the gray circles.
The number to the left and right of the circles
show the node indices and child node indices, respectively. The red line connects the leaf nodes whose keys appear in the
corresponding cornerstone array depicted in the lower panel of Fig.\ \ref{fig:cstone_array}.}
\label{fig:octree}
\end{figure}
In Fig.\ \ref{fig:octree} we illustrate this format with an example. The gray circles correpond to the nodes and contain
the SFC keys from $\mathbf{NK}$, each of which start with a 1-digit (the placeholder bit) followed by $l$ octal digits where
$l$ is the tree level of the node. The numbers to the left of the circles indicate the node indices and the numbers to the right
show the values of the $\mathbf{CO}$ array. For this example, the tree level offsets are
$\mathbf{LO} = \lbrace 0, 1, 9, 17, \ldots, 17 \rbrace$.

To generate octrees in this format, we perform the following steps:
\begin{enumerate}
    \item Copy $\mathbf{K}$ to $\mathbf{NK}[n_i,n]$ and convert to the Warren-Salmon placeholder-bit format \label{item:leafcopy}
    \item Generate internal node keys $\mathbf{NK}[0, n_i]$:
          For elements of $\mathbf{K}$ in parallel compute \texttt{d = clz(K[i] \^\ K[i+1])},
          where \texttt{clz} counts the number of leading zero bits. If $ \mathrm{mod}(d, 3) = 0$,
          $\mathbf{K}_i$ is the key of an internal node at level $d/3$. The corresponding placeholder-bit key
          is stored at $\mathbf{NK}_j$ with $j = (i - \delta(\mathbf{K}_i, d)) / 7$, where $\delta$ is defined below.
          This step can be combined with step \ref{item:leafcopy}.
    \item Radix sort $\mathbf{NK}$ to yield a breadth-first node ordering.
    \item Determine $\mathbf{LO}$ with a binary-search over $\mathbf{NK}$.
    \item Locate the first child of internal nodes with a binary-search,
          using $\mathbf{LO}$ to narrow the search range and store in $\mathbf{CO}$.
\end{enumerate}
The $\delta$ function that maps indices of leaf nodes to indices of internal nodes is given by
\begin{equation}
    \delta(\mathbf{K}_i, d) = \sum_l^{d/3+1} \mu_{k_l}, \quad \mu_0, \ldots, \mu_7 = \lbrace 0, -1, -2, -3, 3, 2, 1, 0 \rbrace,
       \label{eq:digitweight}
\end{equation}
where $\mathbf{K}_i = k_1 k_2 \ldots k_L$, such that $k_l$ corresponds to the $l$-th octal digit of $\mathbf{K}_i$, counting from the most
significant digit. The resulting ordering of internal octree nodes before the sorting step is equal to what one would get by first constructing
a binary radix tree on top of $\mathbf{K}$ according to Karras, followed by enumeration and extraction of the octree nodes with a prefix sum.
Indeed, we deduced Eq.\ \eqref{eq:digitweight} from the layout of the binary radix tree by induction under the assumption that
each internal node has 8 children.

Additional octree node properties, such as particle counts, expansion centers, multipole moments, etc can be stored in separate arrays of length $n$
and may be constructed in subsequent steps when required. We choose a breadth-first ordering of the octree nodes to ensure that the child nodes
of each cell are always next to each other and to allow for efficient traversal in Barnes-Hut treecodes \cite{Barnes1990, Bonsai2012}.

Finally, to provide an overview of the relative costs of the steps involved in the tree construction,
we list in Table\ \ref{tab:treetime} timings for 1, 32 and 512 million particles resulting in octrees
with $55800$, $1.78$ million and $28.9$ million leaf nodes, respectively.
\begin{table}
\caption{Timings for tree operations on an AMD EPYC 7713 64-core CPU, MI250X GPU and an NVIDIA A100 GPU.
         Either 1, 32 or 512 million particles were distributed with a 3D Gaussian density profile and $N_{crit}$
         was set to $64$. The resulting octrees have $55.8$K, $1.78$M and $28.9$M leaves, respectively.
         The radix sort step was implemented with \texttt{sort\_by\_key} from the Thrust library on both the CPU and GPU,
         with the particle SFC keys provided as keys and the identity permutation as values to keep track of the ordering.}
\label{tab:treetime}
\begin{tabular}{lccc}
\toprule
\textit{execution on CPU}                  & 1M  & 32M &        512M\\
\midrule                                                            
64-bit Morton keys                         &  2.2 ms  & 33  ms & 520 ms \\
64-bit Hilbert keys                        &  3.2 ms  & 107 ms & 1640 ms\\
Radix sort                                 &  23 ms   & 607 ms & 13100 ms\\
Leaf update (Sec. \ref{sec:leaftree})      &  0.47 ms & 32  ms & 151 ms\\
Internal nodes (steps 1-5)                 &  2.6 ms  & 80  ms & 1400 ms\\
\midrule
\textit{execution on MI250X}               & 1M        & 32M   & 512M \\
\midrule                                                            
64-bit Morton keys                         & 0.041 ms  & 0.81 ms & 13 ms\\
64-bit Hilbert keys                        & 4.3   ms  & 7.3  ms & 43 ms\\
Radix sort                                 & 0.925 ms  & 22  ms  &  330 ms\\
Leaf update (Sec. \ref{sec:leaftree})      & 0.14  ms  & 0.81 ms & 5.8 ms\\
Internal nodes (steps 1-5)                 & 4.9   ms  & 6.3 ms  &  24 ms\\
\midrule
\textit{execution on A100}                  & 1M        & 32M   & 512M \\
\midrule                                                            
64-bit Morton keys                         & 0.018 ms  & 0.31 ms &  7.9 ms\\
64-bit Hilbert keys                        & 0.084 ms  & 1.9  ms &  30 ms\\
Radix sort                                 & 0.483 ms  & 5.0  ms &  73 ms\\
Leaf update (Sec. \ref{sec:leaftree})      & 0.11  ms  & 0.51 ms &  2.5 ms\\
Internal nodes (steps 1-5)                 & 0.19  ms  & 0.65 ms &  5.4 ms\\
\bottomrule
\end{tabular}
\end{table}

\section{Distributed octrees and domain decomposition}
\label{sec:distributedtree}

In the case where particles and their sorted SFC keys $\mathbf{P}$ are distributed across
multiple compute nodes, we have to extend the algorithm in
Sec.\ \ref{sec:leaftree} with a communication scheme. With a small modification,
we can identically replicate across compute nodes the octree that arises from the globally unified set of particles.
To that end, after applying $f$,
we perform an element-wise global sum of the histogram counts array $\mathbf{N}$:
\begin{eqnarray}
 f_{\mathrm{glob}} &= f_r \circ f,
\end{eqnarray}
where $f_r$ simply computes an element-wise global sum of $\mathbf{N}$.
The additional reduction step $f_r$ can be implemented with a single call to \textit{MPI\_Allreduce},
supplying $\mathbf{N}$ as argument. Initializing $\mathbf{K}$ to the root node on all compute nodes, followed by an
alternating application of $f_{\mathrm{global}}$ and $g$ until converged, then yields
the balanced cornerstone array, replicated on all compute nodes, for the distributed set of SFC keys $\mathbf{P}$.
Due to the global collective communication pattern of $f_r$, this method is not suitable for generating octrees
with a high resolution. However, it is fast enough for generating a global octree with sufficient
resolution for decomposing the global domain by partitioning the space filling curve.
The global domain can then be partitioned into $n_\mathrm{ranks}$ subdomains by grouping $\mathbf{K}$ into
$n_\mathrm{ranks}$ bins of consecutive leaves such that each bin contains approximately $N_\mathrm{tot}/n_\mathrm{ranks}$ of
the total particle count. Each bin covers a unique SFC key range that defines a compact volume of fractal-like shape.
Our domain decomposition based on a coarse globally replicated octree requires collective
communication that scales as $\mathcal{O}(N_\mathrm{tot} \log{N_\mathrm{tot}})$ and
presents an alternative to other methods, such as sampling of the SFC \cite{Blackston1997, Bonsai2014} which scales as
$\mathcal{O}(N_\mathrm{tot}^2)$ \cite{Speck2011}.

\section{Locally essential octrees}
\label{sec:essentialtree}

\subsection{Definition}

Let us assume that we have decomposed the global domain into $n_\mathrm{ranks}$ subdomains as described
in Sec.\ \ref{sec:distributedtree}.
In order for the domain decomposition to balance the subdomain particle counts roughly within $p\%$ of
$N_\mathrm{tot}/n_\mathrm{ranks}$-th where $N_\mathrm{tot}$ is the total global number of particles,
only about $100 / p$ leaf nodes are needed per subdomain.
Consequently, the global tree will have a leaf node count resolution of
$N_\mathrm{crit} = \frac{N_\mathrm{tot} \cdot p}{n_\mathrm{ranks} \cdot 100}$
which is far too coarse for efficiently calculating physical forces for particles within a subdomain.
What would be the requirements of a LET that facilitates the calculation of forces
within just one subdomain, containing only the minimum number of nodes required for that task?
In order to address this question, we need to introduce some physical context.

Two important types of forces covering a wide range of physical applications are
\begin{itemize}
  \item Short-range forces with a cutoff. Examples: Smoothed Particle Hydrodynamics (SPH),
    Van-der-Waals dispersion in molecular dynamics
  \item Long-range forces arising from $1/r$ potentials. Examples: gravity and electrostatics
\end{itemize}

For the calculation of short-range forces, the octree has to have a high resolution (small $N_\mathrm{crit}$)
inside a given subdomain $F$ for accurate neighbor particle searches. Additionally, a high resolution
is also required close to the exterior surface for the discovery of halo particles.

Concerning long-range forces, we assume that the algorithm used to calculate them is either the Barnes-Hut treecode or FMM.
Inside and close to the exterior surface of $F$, the tree resolution requirements are identical to
the short-range case. Outside $F$ however, a geometrical criterion, commonly called Multipole Acceptance Criterion (MAC),
applies. For a given source tree node and target point or node, the MAC describes whether the application of the
multipole approximation of the force due to the source on the target is acceptable or not.
What all MACs have in common is that the largest acceptable size of a source node correlates with the opening angle,
i.e. the ratio between source node size and distance to the target.
One of the simplest possible MAC choice is the minimum distance MAC, which is fulfilled if
\begin{equation}
  r_\mathrm{min} > l / \theta,
\end{equation}
where $r_\mathrm{min}$ is the minimum distance between source and target, $l$ is the source node edge length and
$\theta$ is an accuracy parameter. The smaller the value for $\theta$ becomes, the smaller the error of the resulting force.
More sophisticated acceptance criteria that depend on additional node properties exist and are compatible with our
definition of an LET that follows.

Let $F$ be a subdomain covering a contiguous range of the space filling curve.
An octree, represented by the cornerstone array $\mathbf{K}^\mathrm{foc}$, is locally essential in $F$ or focused on $F$ if
\begin{enumerate}
  \item Leaves in $F$ have a particle count $\leq N_\mathrm{crit}$ while the counts of internal nodes in $F$
        exceed $N_\mathrm{crit}$, i.e. they are balanced per the definition in Sec.\ \ref{sec:leaftree}. \label{item:infocus}

  \item Leaves outside $F$ either pass the chosen MAC with respect to any point in $F$ or their particle count
        is smaller than $N_\mathrm{crit}$. The motivation for this condition is to make sure that when Barnes-Hut treecode
        or FMM is applied to particles in $F$, if tree traversal encounters a leaf node outside $F$, either the multipole
        approximation is valid or it contains only a small number of particles that can be transferred at a small cost.
        Consequently, refinement of nodes at the exterior surface of $F$ stops once particle counts drop below $N_\mathrm{crit}$
        as illustrated in Fig.\ \ref{fig:quadtree}.
        \label{item:endpoints}

  \item Internal nodes outside $F$ fail the chosen MAC for at least one point or node in $F$ and their particle
        count is $> N_\mathrm{crit}$. This ensures that
        the local tree depth is the minimum required to satisfy condition\ \ref{item:endpoints}.
\end{enumerate}
If we assume that particles within $F$ reside on a particular compute node that additionally
holds a copy of the halo particles close to the surface, the definition of the LET ensures
that we will be able to compute long-range forces for particles in $F$ depending on global information without the
need for access to particle data beyond $F$ and its surface.
Insofar, it is important to note that the LET still covers the entire global domain.

\subsection{Construction}

As was the case for local or globally replicated octrees, we represent a LET with a cornerstone array.
The general iterative construction scheme that yields an octree that satisfies the LET definition
also remains largely the same: for each leaf node, the decision to either subdivide,
keep, or merge it is encoded in the same fashion, i.e. functions $g_2$ and $g_3$ apply without modification.
What is different is that apart from the node counts array $\mathbf{N}^\mathrm{foc}$, the rebalancing operations to be performed
now depend on a second array $\mathbf{M}$ containing MAC evaluations and that the determination of $\mathbf{N}^\mathrm{foc}$
requires a combination of collective and point-to-point communication.
The MAC evaluations stored in array $\mathbf{M}$ are performed by a function denoted $f^\mathrm{LE}_2$
which traverses the octree defined by $\mathbf{K}^\mathrm{foc}$.
The result of $f^\mathrm{LE}_2$ are boolean values for each octree node outside $F$ that describes whether
there exists a leaf in $F$ combined with which the MAC fails. If one chooses the minimum distance MAC, the
evaluation only depends on the geometrical size and location of the tree nodes and thus does not require communication.
More adaptive MAC variants may rely on additional tree node properties, however, such as multipole expansion centers,
which have to be computed prior to evaluating the MACs and will require communication.
As the computation of $\mathbf{M}$ involves tree traversal, $f^\mathrm{LE}_2$ constructs the fully linked octree
that matches the leaf nodes $\mathbf{K}^\mathrm{foc}$ according to the procedure described in Sec.~\ref{sec:linkedtree}.

An update cycle that refines the LET from $\mathbf{K}^\mathrm{foc}$ to $\mathbf{K'}^\mathrm{foc}$
consists of the following steps:
\begin{eqnarray}
    &&f^\mathrm{LE}_1: (\mathbf{P}, \mathbf{K}^\mathrm{foc}, \mathbf{K}^\mathrm{glob}, \mathbf{N}^\mathrm{glob})
            \mapsto \mathbf{N\phantom{'}} \quad (\text{particle counts}) \label{eq:histogramf}\\
    &&f^\mathrm{LE}_2: \mathbf{K}^\mathrm{foc} \mapsto \mathbf{M\phantom{'}} \quad (\text{MAC evaluations}) \label{eq:mac}\\
    &&g^\mathrm{LE}: (\mathbf{K}^\mathrm{foc}, \mathbf{N}^\mathrm{foc}, \mathbf{M}) \mapsto \mathbf{K'}^\mathrm{foc}, \quad (\text{rebalancing}) \label{eq:rebalancef}
\end{eqnarray}
where $\mathbf{K}^\mathrm{glob}$ is the cornerstone array of the globally replicated octree used for domain decomposition.
We again distinguish between functions labelled $f$ which compute node properties without modifying the octree, such as $\mathbf{N}$ and $\mathbf{M}$, and functions labelled $g$ which rebalance the octree based on said properties.
A further distinction between the two groups is that the computation of node properties requires communication,
while rebalancing consists of local operations only. The function $g^\mathrm{LE}$ decomposes into steps
$g^\mathrm{LE} = g^\mathrm{LE}_1 \circ g_2 \circ g_3$ with
\begin{equation}
    g^\mathrm{LE}_1: (\mathbf{K}^\mathrm{foc}, \mathbf{N}^\mathrm{foc}, \mathbf{M}) \mapsto \mathbf{O}, \quad (\text{rebalancing decision})
\end{equation}
and $g_2$ and $g_3$ as defined in Sec.\ \ref{sec:leaftree}. For nodes inside $F$, $g^\mathrm{LE}_1$ is identical to
$g_1$, i.e. only particle counts are taken into account to compute the corresponding elements of $\mathbf{O}$.
Outside $F$, the result of $g^\mathrm{LE}_1$ depends on the particle counts as well as the MAC evaluations:
a node is only split if the particle count exceeds $N_{crit}$ and also the MAC evaluation failed.
Conversely, a node is merged if its parent node either has a particle count smaller than $N_{crit}$ or
if the parent passed the MAC evaluation.

\begin{figure}
\includegraphics[width=\linewidth]{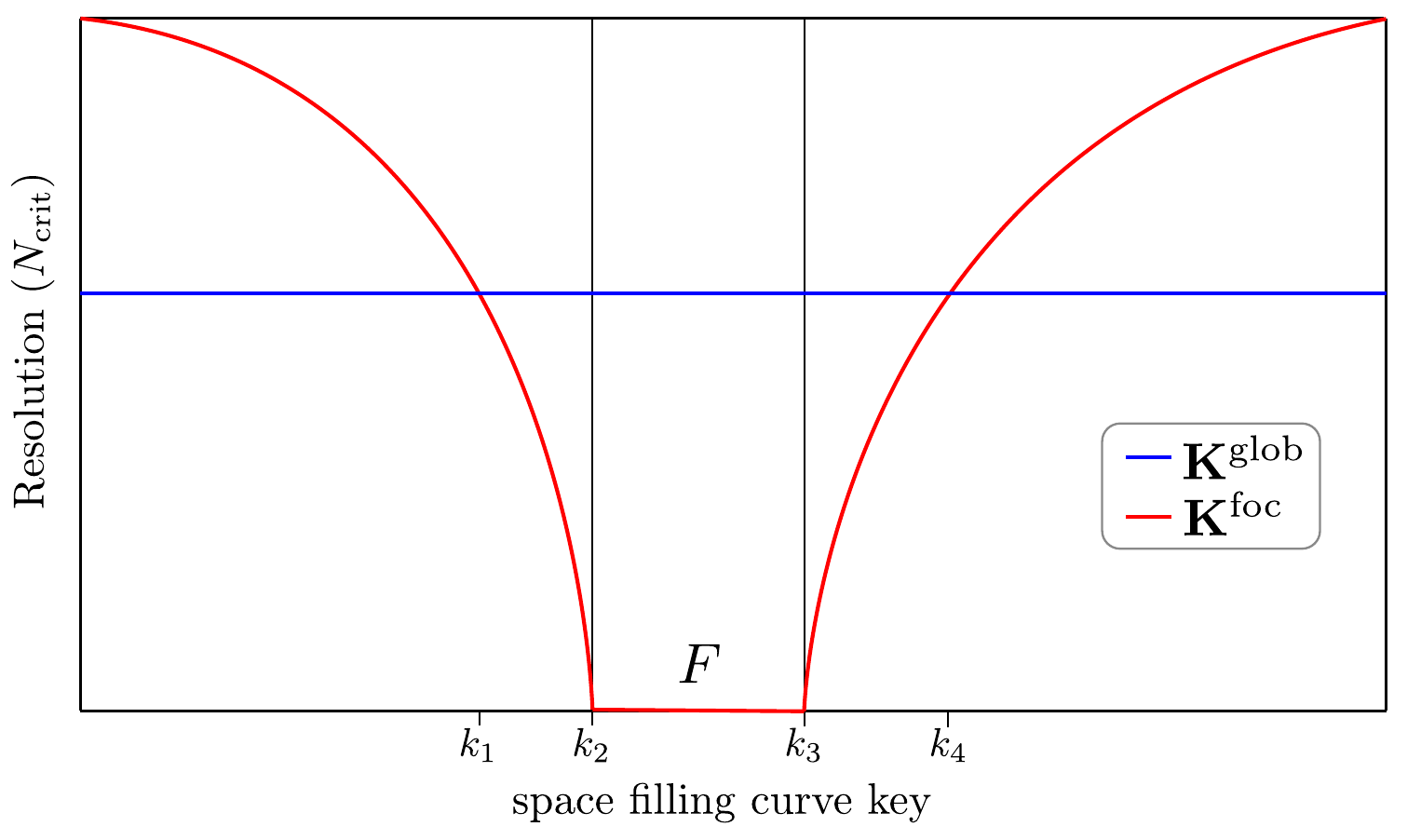}
\caption{Resolution in terms of $N_\mathrm{crit}$ of the global octree and LET along the space filling curve.
         The line for $\mathbf{K}^\mathrm{foc}$ outside $F$ is drawn schematically. If plotted true to scale, $\mathbf{K}^\mathrm{foc}$
         would fall below $\mathbf{K}^\mathrm{glob}$ in several areas, not just to the left and right of $F$.
         Note that the straight line for the global octree does not imply a regular grid structure; octree nodes of varying sizes manifest
         in this plot as stretches of different lengths along the $x$-axis.}
\label{fig:focustree}
\end{figure}

The most demanding function to implement is $f^\mathrm{LE}_1$.
It calculates node properties of the LET which can be computed by an upward pass.
This applies to properties where the leaves depend on the contained particles and internal nodes on their children.
Examples are given by node particle counts $\mathbf{N}$, multipole expansion centers, i.e. center-of-mass for gravity, and multipole moments.
In the following, we assume that the domain decomposition was carried out based on $\mathbf{K}^\mathrm{glob}$, assigning to each MPI rank
a subdomain $F$ along with the particles contained therein.
The aim of $f^\mathrm{LE}_1$ is now to compute an array $\mathbf{Q}$ of node properties arranged in the same layout as the nodes
of the fully linked octree that can be constructed on top of $\mathbf{K}^\mathrm{foc}$. 
Our strategy is to first construct the elements of
$\mathbf{Q}$ that correspond to the leaf nodes as this allows each rank to obtain the internal part through an upward pass without
any further communication. The elements of $\mathbf{Q}$ that correspond to leaves in $\mathbf{K}^\mathrm{foc}$ fall into three
different categories which are illustrated in Fig.\ \ref{fig:focustree}:
\begin{enumerate}
    \item inside $F$ between keys $k_2$ and $k_3$
    \item outside $F$ and exceeding the resolution of the global octree between keys $k_1$ to $k_2$ and $k_3$ to $k_4$.
    \item outside $F$ and contained in $\mathbf{K}^\mathrm{glob}$ to the left of $k_1$ and to the right of $k_4$.
\end{enumerate}
While properties of leaves inside $F$ can be directly computed based on local particle data, the second category involves
point-to-point communication with nearby subdomains, and the third category involves collective communication followed by an upward pass
of the global octree.
Before any communication can take place, each rank must first build the fully linked octree on top of $\mathbf{K}^\mathrm{foc}$, compute the elements
of $\mathbf{Q}$ tied to leaves in $F$ and perform an upward pass. This yields a partially filled property array $\mathbf{Q}$ containing
valid data for any node contained in $F$. Additionally, each rank determines a list of peer ranks that contain category $2$ tree nodes in
their subdomains by traversing the global octree.
Point-to-point communication and collective communication may then proceed in parallel.
The former is performed in two steps: in a first round of point-to-point messages, each rank sends a request to its peers that defines
a series of octree nodes. As only leaf nodes are requested, the messages contain sub-sequences of $\mathbf{K}^\mathrm{foc}$.
On the receiving side, the transmitted octree nodes may correspond to either leaf or internal nodes of the LET.
An answer is then assembled by locating each requested octree node in the LET of the receiver and extracting
the corresponding elements of $\mathbf{Q}$. The point-to-point exchange concludes with a second round of messages where the peer ranks
send the required properties back to the requester.
In order to obtain the remaining properties of nodes that fall into the third category, we construct an auxilliary array of properties
$\mathbf{Q}^\mathrm{glob}$ whose node layout matches the fully linked octree constructed from $\mathbf{K}^\mathrm{glob}$.
Each rank has the required data to populate the leaf nodes of $\mathbf{Q}^\mathrm{glob}$ that fall into its subdomain.
We can then perform an \textit{MPI\_Allgatherv} operation on the $\mathbf{Q}^\mathrm{glob}$ array to replicate the leaf node quantities on all ranks.
A subsequent upward pass of $\mathbf{Q}^\mathrm{glob}$ yields the internal nodes up to the global root node.
Finally, each rank can complete the leaves of $\mathbf{Q}$ by extracting the missing nodes from $\mathbf{Q}^\mathrm{glob}$.
Since some internal nodes in the LET may exist outside $F$ and also not appear in the global octree, a second
upsweep of $\mathbf{Q}$ is needed to complete its construction. The final properties may then serve as the basis for particle force
computations and as criteria to rebalance $\mathbf{K}^\mathrm{foc}$ in Eq.\ \eqref{eq:rebalancef}.

\begin{figure}
\includegraphics[width=0.8\linewidth]{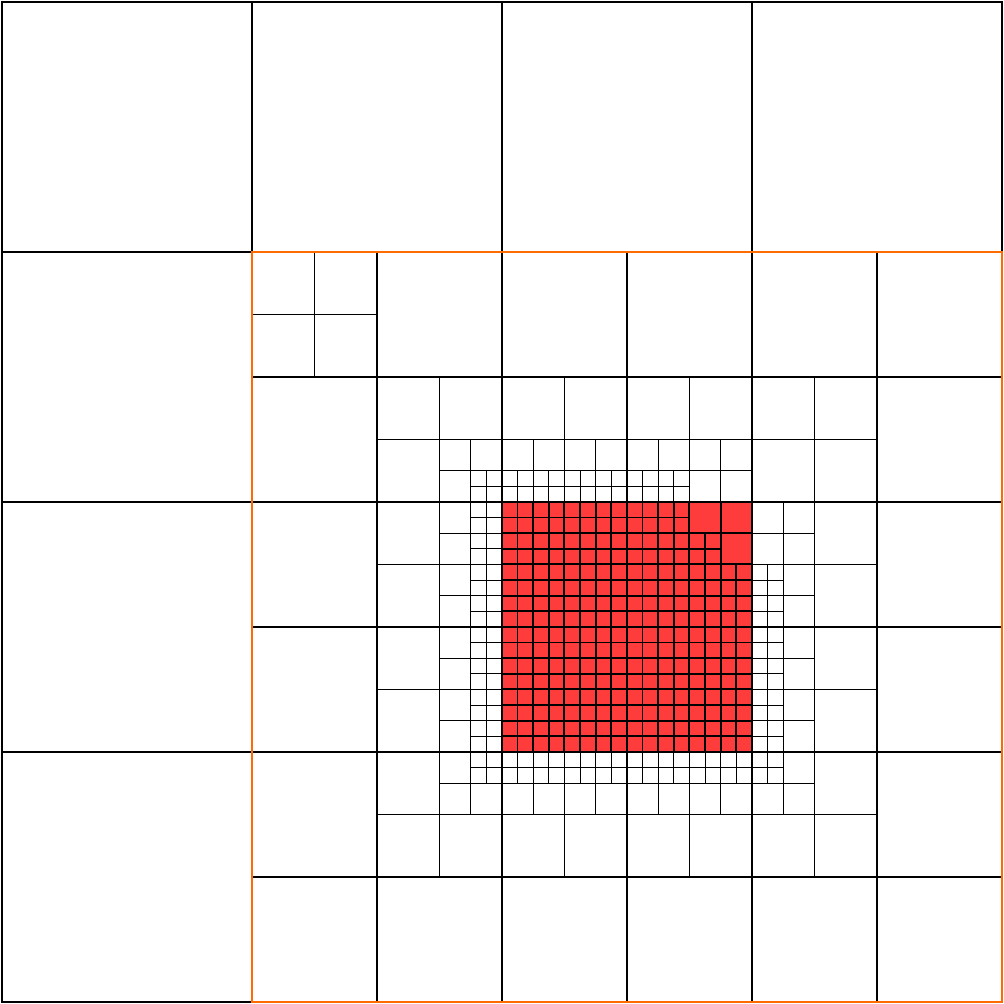}
\caption{Example of a locally essential quadtree with the focus area $F$ highlighed in red.
         The orange frame encompasses the part of $\mathbf{K}^\mathrm{foc}$ where the
         resolution exceeds $\mathbf{K}^\mathrm{glob}$. }
\label{fig:quadtree}
\end{figure}

We implement all stages of LET generation on GPUs. In addition to the operations discussed in Sec.\ \ref{sec:leaftree}, this involves
and upsweep in $f^\mathrm{LE}_2$ and evaluating MACs in $f^\mathrm{LE}_2$, for which we employ fully linked octrees constructed on
top of $\mathbf{K}^\mathrm{glob}$ and $\mathbf{K}^\mathrm{foc}$.
After the application of Eqs.\ \eqref{eq:histogramf} to \eqref{eq:rebalancef} no longer result in any changes to the LET,
an analogous locally essential quadtree might look as shown in Fig.\ \ref{fig:quadtree}.
In general, the resolution will be highest in the focus area $F$ highlighted in red
and decrease when moving away from $F$. But since the local particle density is also taken into account, the tree structure
does not necessarily follow a uniform geometrical pattern.
In our construction scheme,
both $\mathbf{K}^\mathrm{glob}$ and $\mathbf{K}^\mathrm{foc}$ are stored in cornerstone format which
is also used exclusively as the format for communication. Fully linked octrees are constructed on-demand
on top of $\mathbf{K}^\mathrm{glob}$ and $\mathbf{K}^\mathrm{foc}$. They are not preserved between time-steps.
The collective communication that we perform on $\mathbf{Q}^\mathrm{glob}$ scales as $\mathcal{O}(n_\mathrm{ranks} \log n_\mathrm{ranks})$
if the number of particles per MPI rank is kept constant, which is worse than the theoretical optimum of $\mathcal{O}(\log n_\mathrm{ranks})$.
Due to the dynamically changing fractal sub-domain geometries that result from dividing the SFC into segments,
a strictly hierarchical $\mathcal{O}(\log n_\mathrm{ranks})$ communication scheme that does not resort to collective communication
for the highest levels of the LET where nodes transcend the sub-domain boundaries is very difficult to implement.
This fact was also pointed out in Ref.\ \cite{Bonsai2014}, and, to the best of our knowledge, remains an unsolved problem.

\section{Results}
\label{sec:results}

\begin{figure}
\includegraphics[width=\linewidth]{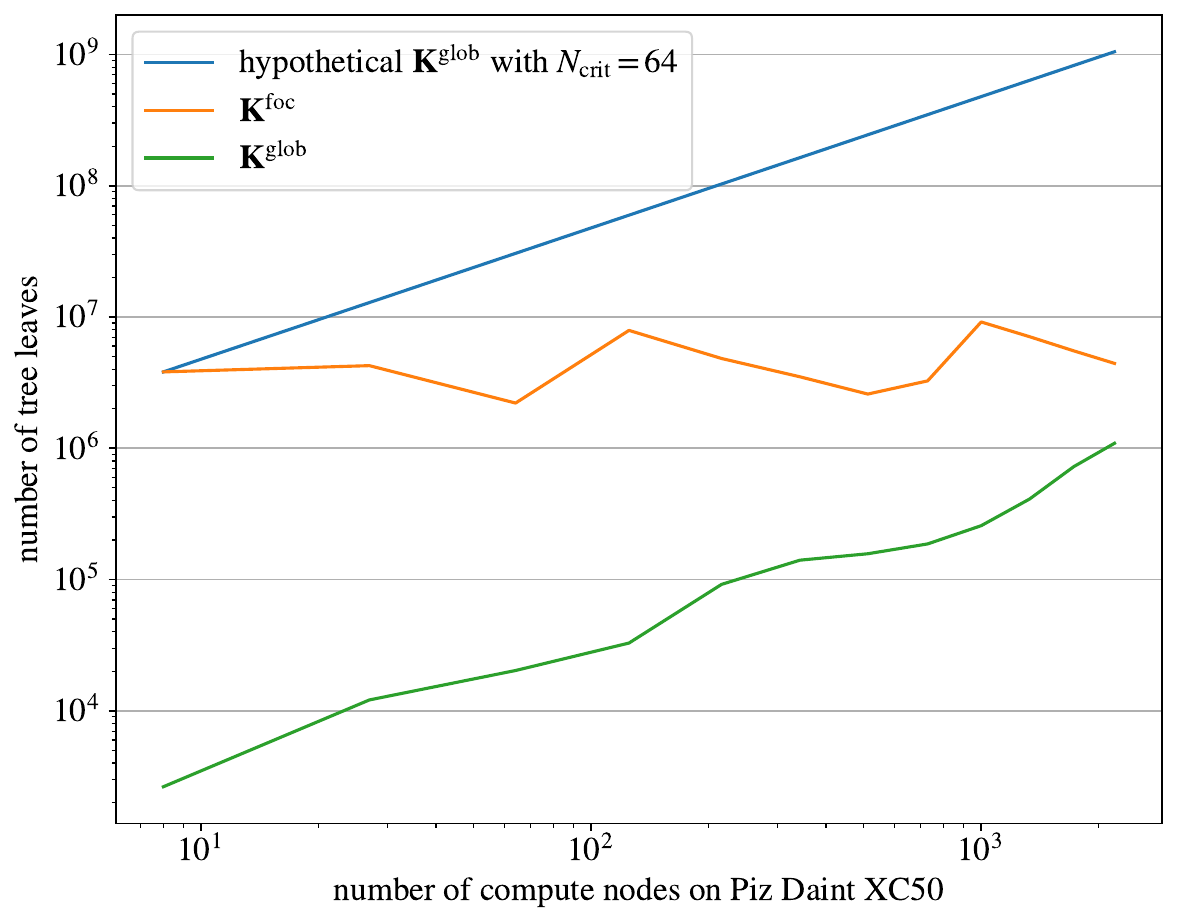}
\caption{Scaling of the global octrees and LETs with system size between 64 million and 141 billion particles.
         The number of particles per compute node was kept constant at 64 million.}
\label{fig:leafcounts}
\end{figure}

To assess the impact of the LET on the required size of the global octree,
we create a spherical particle distribution with density inversely proportional to the radius and distribute it
on varying numbers of compute nodes each running one MPI rank assigned with one subdomain constructed according to the
method described in Sec.\ \ref{sec:distributedtree}. The radius of the sphere is kept constant and each compute node
holds a constant count of 64 million particles. With increasing number of compute nodes and total number of particles,
the density of the sphere increases while the volumes of the subdomains decrease. For system sizes between 64 million and 141 billion
particles, we construct LETs with $N_\mathrm{crit} = 64$ and compare the resulting number of
tree leaves against the leaf counts of $\mathbf{K}^\mathrm{glob}$ (i.e. the global octree used for domain decomposition) and
against a hypothetical octree with a uniform resolution of $N_\mathrm{crit} = 64$ everywhere.
In Fig.\ \ref{fig:leafcounts}, we observe that LET growth is consistent with the expected $\mathcal{O}(\log n_\mathrm{ranks})$ behavior.
By construction, $\mathbf{K}^\mathrm{glob}$ grows linearly and will eventually exceed the size of $\mathbf{K}^\mathrm{foc}$,
but it is roughly three orders of magnitude smaller compared to the octree with high uniform resolution that would be required
in the absence of a LET.

\begin{figure}
\includegraphics[width=\linewidth]{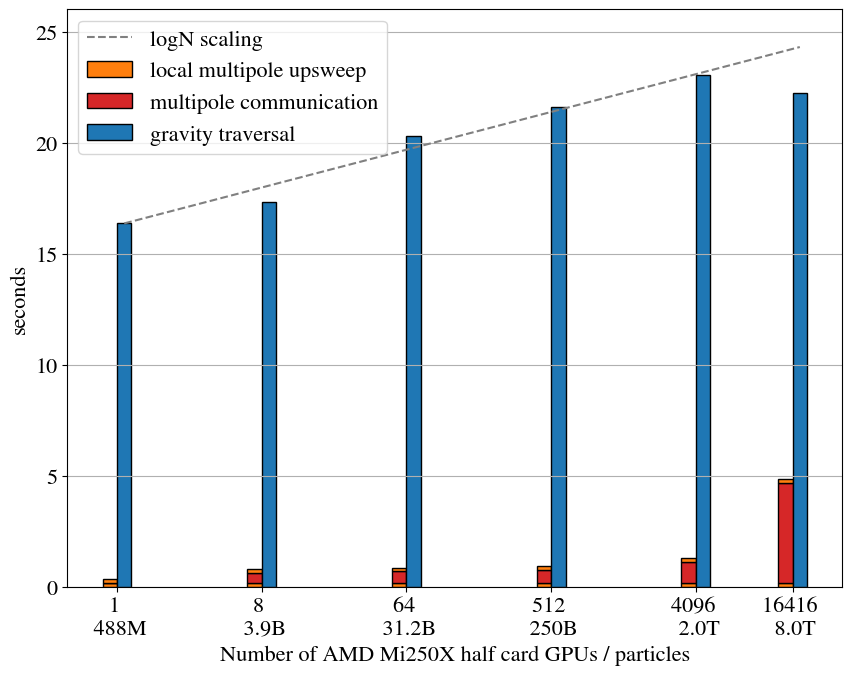}
\caption{Barnes-Hut weak scaling on LUMI-G with 488 million particles per GPU and 8 trillion particles across
         2052 computing nodes. The blue bars show the time required for traversing the LET to compute accelerations.
         Timings for multipole construction are subdivided into local compute and communication parts. Floating point operations
         were carried out in double precision.}
\label{fig:nbody}
\end{figure}

In order to compare the cost of communication and computation, we implement a highly optimized version of Barnes-Hut
treecode and compare the time required for the tree traversal with the the time it takes to compute the multipole moments
for the LET by applying $f^\mathrm{LE}_1$. We used the treecode available at \url{https://github.com/exafmm/bonsai} as
a starting point, a modified version of the original Bonsai code \cite{Bonsai2012} with breadth-first traversal for GPUs.
We further modified this code by changing the octree structure to our LET, replacing PTX inline assembly instructions
from the NVIDIA Kepler microarchitecture with portable warp-level instructions for compatibility with AMD GPUs,
and by adding further optimizations. In Fig.\ \ref{fig:nbody}, we test the performance and scalability of our Barnes-Hut
implementation on the LUMI-G system that features compute nodes with 4 AMD Instinct MI250X GPU cards. Each card contains
two chips that appear as two separate GPUs and as a consequence, we ran our tests with 8 MPI ranks per compute node.

\begin{figure}
\includegraphics[width=\linewidth]{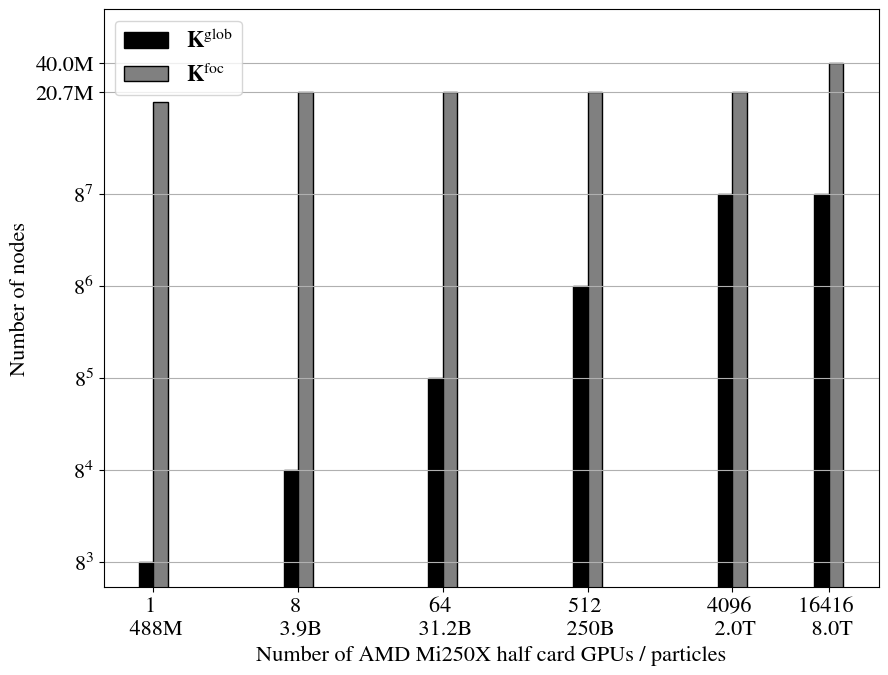}
\caption{Scaling of LET and global octree node counts with system size.}
\label{fig:nbody-tree}
\end{figure}

Our choice of cartesian multipoles with quadrupole corrections and $\theta = 0.5$ results in three significant digits in the computed
particle accelerations. As described in Sec.\ref{sec:essentialtree}, the LET multipole computation consists of two local upsweeps
with a communication phase in between. After the first local upsweep is complete, it is possible to launch the tree traversal starting
from the top LET nodes spanning the local sub-domain in order to achieve overlap with the multipole communication phase.
Once the traversal of the local nodes completes, the remaining LET parts can be traversed by providing the LET nodes to the left and right
of the local sub-domain on the SFC as starting points. For obtaining the results shown in Fig.\ \ref{fig:nbody}, we did not overlap
multipole construction with tree traversal and instead imposed a synchronization barrier in between to properly separate the timings
for the two operations recorded as maxima across ranks. To estimate achieved performance in terms of Flops,
we collected interaction statistics on one rank, counting the average number of particle-particle (P2P) and multipole-particle (M2P)
interactions accrued during traversal per particle. In accordance with Ref.\ \cite{Bonsai2014}, counting P2P and
M2P interactions as 23 and 65 flops, respectively, we estimate the performance per rank and MI250X half card at 6.7 Tflops
in double precision, reaching a peak performance of 110 Pflops for the largest run with
8 trillion particles on 2052 nodes or 16416 GPU half cards, the maximum we were able to use.
Note that the jump in communication time
for the exchange of multipole moments when moving from 2 to 8 trillion particles is not due to an increase in collective communication.
As shown in Fig.\ \ref{fig:nbody-tree}, the number of nodes in the global octree is in fact identical between the last two data points.
The difference in communication time is caused by the larger number of LET nodes in the largest run. This is simply a consequence
of the average number of particles per leaf node in the LET fluctuating between $\approx N_\mathrm{crit}/8$ and $N_\mathrm{crit}$ (64), explaining
the shorter traversal time on 16416 GPUs as well. While the average number of particles per leaf was around 30 on 1 to 4096 GPUs,
the average dropped to 12 for the system with 8 trillion particles,
resulting in higher traversal efficiency by reducing the number of P2P interactions per particle.

\section{Summary and future work}

We have shown that octrees can be efficiently constructed on GPUs with the presented algorithms.
The resulting octrees and LETs are suited to supporting the computation of short and long-ranged forces acting on particles
and can be constructed at scale on distributed HPC systems. Relative to a competitive Barnes-Hut treecode implementation,
LET construction overhead remains small enough to allow for nearly complete overlap with the tree traversal phase
on almost the full LUMI-G HPC system.
Our future goal is to employ Cornerstone for an interesting scientific application,
for example by combining our N-body solver with SPH.

\subsection{Software availability}

The algorithms described in this work have been implemented in C++, CUDA and HIP, and are available as open-source
software. Octrees, LETs, neighbor searching and domain decomposition are bundled into the Cornerstone library available
at \url{https://github.com/sekelle/cornerstone-octree}. It does not implement any physics, such as Barnes-Hut or FMM for gravity.
Cornerstone is integrated into a full-fledged application framework called SPH-EXA,
available at \url{https://github.com/unibas-dmi-hpc/SPH-EXA}. SPH-EXA can perform I/O, generate initial conditions and
offers SPH and gravity (described in Sec.\ \ref{sec:results}) as composable components for physical simulations.
Efforts are underway to also make our Barnes-Hut implementation available as a stand-alone package.

\begin{acks}
This work was supported by the Swiss Platform for Advanced Scientific Computing (PASC) project SPH-EXA (funding periods 2017-2021 and 2021-2024).
We acknowledge the support of the LUMI-G pilot program hosted by CSC, in particular Emmanuel Ory, Pekka Manninen and Fredrik Roberts\'en,
and the Swiss National Supercomputing Center (CSCS) for providing access to the Piz Daint supercomputer.
\end{acks}

\bibliographystyle{ACM-Reference-Format}
\bibliography{refs}

\appendix

\section{Tables of symbols}

\begin{table}[H]
\caption{Symbols introduced in Sec.\ \ref{sec:leaftree} involved in the construction of octree leaves.}
\label{tab:treetime}
\begin{tabular}{cll}
\toprule
\textit{symbol}         & \textit{description}              \\
\midrule
 $L$                    & maximum octree depth \\
 $\mathbf{K}$           & cornerstone leaf node SFC key array, Eq.\ \eqref{eq:cstone} \\
 $n_l$                  & number of leaf nodes in $\mathbf{K}$ \\
 $\mathbf{N}$           & leaf node particle counts array  \\
 $N_\mathrm{crit}$      & maximum number of particles per leaf \\
 $\mathbf{P}$           & array of local particle SFC keys \\
 $f$                    & leaf node particle counting function \\
 $g$, $g_{1,2,3}$       & functions to rebalance $\mathbf{K}$ \\
 $\mathbf{O}$           & auxiliary array for node transformation operations \\
\bottomrule
\end{tabular}
\end{table}

\begin{table}[H]
\caption{Symbols introduced in Sec.\ \ref{sec:linkedtree} involved in the construction of fully linked octrees.}
\label{tab:treetime}
\begin{tabular}{cll}
\toprule
\textit{symbol}         & \textit{description}              \\
\midrule
 $\mathbf{NK}$          & array of node keys for internal and leaf nodes \\
 $n_i$                  & number of internal nodes \\
 $n$                    & total number of nodes, $n = n_i + n_l$ \\
 $\mathbf{CO}$          & child connectivity array \\
 $\mathbf{LO}$          & octree level offsets \\
 $\delta$               & mapping between leaf and internal nodes\\
 $d$                    & number of common leading bits of two SFC keys\\
\bottomrule
\end{tabular}
\end{table}

\begin{table}[H]
\caption{Symbols introduced in Sec.\ \ref{sec:distributedtree} involved in the construction of distributed and
         globally replicated octrees.}
\label{tab:treetime}
\begin{tabular}{cll}
\toprule
\textit{symbol}         & \textit{description}              \\
\midrule
 $f_r$                  & \textit{MPI\_Allreduce} \\
 $f_\mathrm{glob}$      & combination of $f$ with $f_r$ \\
 $n_\mathrm{ranks}$     & number of MPI ranks or subdomains\\
 $N_\mathrm{tot}$       & total number of particles across all ranks \\
\bottomrule
\end{tabular}
\end{table}

\begin{table}[H]
\caption{Symbols introduced in Sec.\ \ref{sec:essentialtree} involved in the construction of focused octrees, also known as LETs.}
\label{tab:treetime}
\begin{tabular}{cll}
\toprule
\textit{symbol}         & \textit{description}              \\
\midrule
 $F$                        & a subdomain corresponding to a section of the SFC \\
 $\mathbf{K}^\mathrm{foc}$  & cornerstone array of an octree focused on $F$ \\
 $\mathbf{N}^\mathrm{foc}$  & particle counts of nodes in $\mathbf{K}^\mathrm{foc}$ \\
 $\mathbf{K}^\mathrm{glob}$ & cornerstone array of a globally replicated octree \\
 $\mathbf{N}^\mathrm{glob}$ & particle counts of nodes in $\mathbf{K}^\mathrm{glob}$ \\
 $\theta$                   & angle-based accuracy parameter \\
 $\mathbf{M}$               & array of length $n$ with MAC evaluation results \\
 $f^\mathrm{LE}_1$          & particle counting function for nodes $\mathbf{K}^\mathrm{foc}$\\
 $f^\mathrm{LE}_2$          & MAC evaluation function \\
 $g^\mathrm{LE}$, $g^\mathrm{LE}_{1}$  & functions to rebalance $\mathbf{K}^\mathrm{foc}$ \\
 $\mathbf{Q}$               & generic node properties of a fully linked LET\\
 $\mathbf{Q}^\mathrm{glob}$ & generic node properties of a globally replicated octree\\
\bottomrule
\end{tabular}
\end{table}

\end{document}